\documentclass[%
 aip,
 amsmath,amssymb,
 reprint,%
]{revtex4-1}

\usepackage{aas_macros}%
\usepackage{graphicx}
\usepackage{dcolumn}
\usepackage{bm}

\usepackage[utf8]{inputenc}
\usepackage[T1]{fontenc}
\usepackage{mathptmx}

\begin{document}

\title[Numerical scheme for evaluating triple interactions in relativistic plasma]{Numerical scheme for evaluating the collision integrals for triple interactions in relativistic plasma}

\author{M. A. Prakapenia}
 \email{nikprokopenya@gmail.com}
 \affiliation{ICRANet-Minsk, Institute of physics, National academy of sciences of Belarus, 220072 Nezale\v znasci Av. 68-2, Minsk, Belarus}
 \affiliation{Department of Theoretical Physics and Astrophysics, Belarusian State University, 220030 Nezale\v znasci Av. 4, Minsk, Belarus}
\author{I. A. Siutsou}%
 \email{i.siutsou@dragon.bas-net.by}
\affiliation{Institute of Physics, Belarus National Academy of Sciences, Nezavisimosti Ave.  68, 220072 Minsk, Belarus}

\author{G. V. Vereshchagin}
\email{veresh@icra.it}
\affiliation{ICRANet, 65122 Piazza della Repubblica, 10, Pescara, Italy}
\affiliation{INAF -- Istituto di Astrofisica e Planetologia Spaziali, 00133 Via del Fosso del Cavaliere, 100, Rome, Italy}

\date{\today}

\begin{abstract}
Binary interactions in relativistic plasma, such as Coulomb and Compton scattering as well as pair creation and annihilation are well known and studied in details. Triple interactions, namely relativistic bremsstrahlung, double Compton scattering, radiative pair production, triple pair production/annihilation and their inverse processes, are usually considered as emission processes in astrophysical problems, as well as in laboratory plasmas. Their role in plasma kinetics is fundamental \cite{2007PhRvL..99l5003A}. We present a new conservative scheme for computation of Uehling-Uhlenbeck collision integral for all triple interactions in relativistic plasma based on direct integration of exact QED matrix elements. Reaction rates for thermal distributions are compared, where possible, with the corresponding analytic expressions, showing good agreement. Our results are relevant for quantitative description of relativistic plasmas out of equilibrium, both in astrophysical and laboratory conditions. 
\end{abstract}

\maketitle

\section{Introduction}

Electron-positron plasma is present in many astrophysical systems \cite{2010PhR...487....1R,vereshchagin2017relativistic}. It is also studied in laboratory experiments with ultra-intense lasers and high energy beams \cite{2012RvMP...84.1177D}.
Relativistic plasma has been the subject of extensive research since the 1970s. Many works focused on binary relaxation via Coulomb collisions between ions and electrons \cite{1976PhRvA..13.1563W,1981PhFl...24..102G,1982ApJ...254..755G,1983MNRAS.202..467S,1990MNRAS.245..453C,1985ApJ...295...28D}, with a special role of electron-positron creation and annihilation process \cite{1982ApJ...258..335S,1984ApJ...283..842Z}. Most papers consider an optically thin or mildly optically thick regime \cite{1971SvA....15...17B,1982ApJ...253..842L,1982ApJ...258..335S,1985MNRAS.212..523G}. Equilibrium in the pair plasma has been studied in \cite{1982ApJ...253..842L,1982ApJ...258..335S,1985MNRAS.212..523G}. The energy loss rate by different processes has been analyzed in \cite{1982ApJ...258..321S,1985A&A...148..386H,1988A&A...191..181H}. The conditions of dominance of the double Compton scattering versus bremsstrahlung were discussed in \cite{1981ApJ...244..392L,1984ApJ...285..275G}.

While the binary interactions between photons and electrons are the subject of classical textbooks in QED, the triple interactions such as electron-electron bremsstrahlung, double Compton scattering, three photon annihilation radiative pair production and their inverse are less represented in the literature. The theory of these processes is reviewed in \cite{1969RvMP...41..581M,1974RvMP...46..815T,1976spr..book.....J}.
The differential cross section and energy loss rate of electron-electron bremsstrahlung is studied in \cite{1975ZNatA..30.1546H,1975ZNatA..30.1099H}, and interaction rates are obtained for non-relativistic \cite{1967PhRv..163..156M} and ultra-relativistic cases \cite{1968PhRv..165..253A}, respectively. The case of electron-positron bremsstrahlung is considered in \cite{1985PhRvD..31.2120H}. 
The differential cross section for the radiative pair production is obtained in \cite{1985ZNatA..40.1182H}, while the total cross section is found in \cite{1981ZNatA..36..413H}. The three-photon annihilation is studied in \cite{1949PhRv...75.1963O,2008PhLA..372.6396F}.
The theory of the double Compton scattering is developed in \cite{1952RSPSA.215..497M}. The cross-section for the non-relativistic case is obtained in \cite{1979ApJ...230..967G,1984ApJ...285..275G}, the emission rate in the soft photon limit is obtained in \cite{2007A&A...468..785C}, while the rate at non-relativistic temperatures is derived in \cite{1981ApJ...251..713L,1981MNRAS.194..439T}. In summary, this vast literature presents analytic expressions for reaction rates in equilibrium. Non-equilibrium rates in relativistic regime can be computed only numerically.

Astrophysical observations indicate that out of equilibrium relativistic plasma is present in such sources as active galactic nuclei, binary X-ray sources, microquasars and gamma-ray bursts \cite{vereshchagin2017relativistic}. Energy release in the early Universe may bring primeval relativistic plasma out of equilibrium as well \cite{2020arXiv200511325C}. In laboratory conditions relativistic electron-positron jets are generated by interaction of intense laser pulses with condensed matter \cite{2015NatCo...6.6747S,2011PPCF...53a5009D,2013JPhCS.454a2016M}.
Hence most general description of relativistic plasma dynamics is required, which is given in terms of distribution function, where particle collisions are described by the integrals of differential cross-section (or a matrix element) over the phase space \cite{cercignani2012relativistic,groot1980relativistic}. So far efficient codes were developed which describe only binary collisions without induced emission \cite{2005ApJ...628..857P}, binary collisions with induced emission \cite{2018JCoPh.373..533P}, or binary and triple collisions not far from thermal equilibrium \cite{2012MNRAS.419.1294C}. In this paper we generalize our method for calculation of collision integrals specifically treating triple interactions in relativistic plasma. In addition, the new kinematic approach, which improve the scheme performance, is presented for triple interactions.

This scheme was first introduced in the work \cite{2004ApJ...609..363A} and then applied to the study of thermalization in relativistic plasma of Boltzmann particles \cite{2007PhRvL..99l5003A,2009AIPC.1111..344A,2009PhRvD..79d3008A,2010AIPC.1205...11A}, for the computation of relaxation timescales \cite{2010PhRvE..81d6401A}, and description of electron-positron plasma creation in strong electric fields \cite{BENEDETTI2013206}. Thermalization process was studied taking into account plasma degeneracy in \cite{2019PhLA..383..306P}. In contrast with non-degenerate plasma described by Boltzmann equations, quantum statistics is taken into account by adopting the Uehling-Uhlenbeck equation, which contains additional Pauli blocking and Bose enhancement multipliers \cite{1934PhRv...46..917U,1933PhRv...43..552U}. It was shown that effects of plasma degeneracy lead to interesting new phenomena such as Bose condensation of photons \cite{2019arXiv190804402P}, and avalanche thermalization due to Pauli blocking \cite{2020arXiv200307288P}.

This paper is structured as follows. In Section 2 the relativistic Boltzmann equation with collision integrals for relativistic plasma is introduced.  In Section 3 numerical treatment for calculation of collision integrals for all binary and triple interactions in relativistic plasma is described. In Section 4 numerical results are reported, and compared and contrasted with existing results in the literature. In Section 5 main results are summarized.

\section{Relativistic Boltzmann equation and collision integrals}

In a homogeneous and isotropic electron-positron plasma kinetic equations for distribution functions
$f$ are \cite{vereshchagin2017relativistic}:
\begin{equation}  \label{Boltzmann_class}
\frac{d}{d t}f(\mathbf{p},t)= \sum_{q}\left(
\eta^{q}-\chi^{q}f(\mathbf{p},t)\right) ,
\end{equation}
where the summation index $q$ denotes all processes of interaction between plasma particles, $\eta^{q}$ and $\chi^{q}$ are emission and absorption coefficients correspondingly, $\mathbf{p}$ is particle three-momentum, $t$ is time. The distribution function  is normalized such that the number density is given by $n= \int f(\mathbf{p}) ~d^3p$. We take into account all binary and triple interactions listed in Table~\ref{binpint}. The treatment of binary interactions is covered by many textbooks on relativistic kinetic theory, see e.g. \cite{cercignani2012relativistic,groot1980relativistic}. Instead, triple interactions are not considered. Accounting for such interactions is essential for understanding radiation from plasma, but also for understanding of thermalization process \cite{2007PhRvL..99l5003A,vereshchagin2017relativistic}.

\begin{table}[ptb]
\caption{Particle interactions in relativistic plasma}\label{binpint}
\centering\small
\begin{tabular}{|c|c|}
\hline
Binary interactions & Triple interactions \\ 
\hline
Compton scattering & Double Compton scattering \\ \hline
$e^{\pm}\gamma{\longrightarrow}e^{\pm\prime}\gamma^{\prime}$ & $e^{\pm}\gamma%
{\longleftrightarrow}e^{\pm\prime}\gamma^{\prime}\gamma^{\prime\prime}$ \\
\hline
Coulomb, M\o ller and Bhabha scattering & Bremsstrahlung \\ \hline
$e_{1}^{\pm}e_{2}^{\pm}\longrightarrow e_{1}^{\pm\prime} e_{2}^{\pm\prime}$
& $e^{\pm}_{1} e^{\pm}_{2}{\longleftrightarrow}e^{\pm\prime}_{1}
e^{\pm\prime }_{2}\gamma$ \\
$e^{+} e^{-}\longrightarrow e^{+\prime} e^{-\prime}$ & $e^{+}e^{-}{
\longleftrightarrow}e^{+\prime}e^{-\prime}\gamma$ \\ \hline
Pair production/annihilation & Three-photon pair \\ &production/annihilation \\ \hline
$e^{+} e^{-}\longleftrightarrow\gamma_{1}\gamma_{2}$ & $e^{+}e^{-}{
\longleftrightarrow}\gamma_{1}\gamma_{2}\gamma_{3}$ \\ \hline
& Radiative pair \\ & production/annihilation \\ \hline
& $\gamma_{1}\gamma_{2}{\longleftrightarrow}e^{+}e^{-}\gamma^{\prime}$ \\
& $e^{\pm}\gamma{\longleftrightarrow}e^{\pm\prime}e^{+}e^{-}$ \\ 
\hline
\end{tabular}%
\end{table}

Consider a triple interaction: two incoming particles of kinds $I$ and $II$ in
quantum states $1$ and $2$ produce three outgoing particles of kinds $III$, $IV$, and $V$ in quantum states $3$, $4$,
and $5$. Let momenta of particles before interaction be $\bf{p}_{1}$ and $\mathbf{p}_{2}$, and after interaction be
$\mathbf{p}_{3}$, $\mathbf{p}_{4}$, and $\mathbf{p}_{5}$, correspondingly. This process can be represented symbolically as
\begin{equation}\label{3pdir}
    I_1 + II_2 \longrightarrow III_3 + IV_4 + V_5.
\end{equation}
The inverse process is
\begin{equation}\label{3pinv}
    III_3 + IV_4 + V_5 \longrightarrow I_1 + II_2.
\end{equation}
Energy $\varepsilon$ and momentum conservation gives
\begin{gather}
\varepsilon_{1} + \varepsilon_{2}=\varepsilon_{3} + \varepsilon_{4} +
\varepsilon_{5},\qquad\mathbf p_{1} + \mathbf p_{2}=\mathbf p_{3} + \mathbf p_{4} + \mathbf
p_{5}.
\end{gather}

In the Uehling-Uhlenbeck equation collision integrals for particle $I$ in the
state $1$ is a function of momentum $\mathbf{p}_{1}$\footnote{The quantum state is defined not only by momentum, but also by the spin
projection. For isotropic distributions complete averaging over spins is performed, and quantum states are denoted by momentum implicitly summing over all possible spin states with the use of spin multiplier $g=2s+1$ for massive particles and $g=2(1)$ for massless particles with spin $s>0$
($s=0$). Therefore in this work plasma particles are assumed to be unpolarized.} and time \cite{1934PhRv...46..917U,1933PhRv...43..552U}
\begin{multline}
\label{3pB}
\eta_{I}(\mathbf{p}_{1},t) -\chi_{I}(\mathbf{p}_{1},t)f_{I}(\mathbf{p}_{1},t)= \int
d^{3} \mathbf{p}_{2} d^{3} \mathbf{p}_{3} d^{3} \mathbf{p}_{4} d^{3} \mathbf{p}_{5} \\*
\quad \times\biggl[W_{(3,4,5|1,2)}
f_{III} f_{IV} f_{V} \times \left(1\pm \bar f_{I}\right) \left( 1\pm \bar f_{II} \right)\\
-W_{(1,2|3,4,5)}
f_{I} f_{II} \times \left(1\pm\bar f_{III}\right) \left(1\pm\bar f_{IV}\right)
\left(1\pm\bar f_{V}\right) \biggr],
\end{multline}
where $\bar f_I=h^{3}f_{I}(\mathbf{p}_{1},t)/g_{I}$ and so on, $g$ being the number of degenerate spin states (for our plasma it is 2 for all components), $h$ is Planck's constant, and $W$ are the transition functions. The first term in the square brackets corresponds to emission of particle $I$ in the inverse process~(\ref{3pinv}), while the second term corresponds to absorption of particle $I$ in the direct process~(\ref{3pdir}). This is the general case when all the kinds of incoming and outgoing particles are different.

It is also possible that particles before and after the interaction are of the same kind. In this case collision integrals become more involved. Without loss of generality consider the case when $I=V$. Then two new terms appear in the collision integral of particle $I$ of state $1$: emission coefficient
$\eta_{I}(\mathbf{p}_{1},t)$ in the direct process $I_5 + II_2 \longrightarrow III_3 + IV_4 + I_1$, and absorption
coefficient $\chi_{I}(\mathbf{p}_{1},t) f_{I}(\mathbf{p}_{1},t)$ in the inverse process $III_3 + IV_4 + I_1 \longrightarrow
I_5 + II_2$. Combining all these terms we obtain the collision integral
\begin{multline}  \label{4termsBoltz}
\eta_{I}(\mathbf{p}_{1},t) -\chi_{I}(\mathbf{p}_{1},t)f_{I}(\mathbf{p}_{1},t) = \int
d^{3}\mathbf{p}_{2} d^{3}\mathbf{p}_{3} d^{3}\mathbf{p}_{4} d^{3}\mathbf{p}_{5} \\*
\times\biggl[ - W_{(1,2|3,4,5)} 
f_{I} f_{II} \times \left(1\pm\bar f_{III}\right) \left(1\pm\bar f_{IV}\right) \left(1\pm\bar f_{V}\right)
\\[-10pt] \ \\
+ W_{(3,4,5|1,2)}
f_{III} f_{IV} f_{V} \times \left(1\pm\bar f_{I}\right)
\left(1\pm\bar f_{II}\right)
\\[-10pt] \ \\
+ W_{(5,2|3,4,1)} 
f_{V} f_{II} \times \left(1\pm\bar f_{III}\right) \left(1\pm\bar f_{IV}\right) \left(1\pm\bar f_{I}\right)
\\[-10pt] \ \\ 
- W_{(3,4,1|5,2)} 
f_{III}(\mathbf{p}_{3},t)f_{IV}(\mathbf{p}_{4},t)f_{I}(\mathbf{p}_{1},t)
\times \left(1\pm\bar f_{V}\right) \left(1\pm\bar f_{II}\right) \biggr].
\end{multline}

All triple QED-processes listed in Table~\ref{binpint}, with the only exception of three-photon pair production/annihilation,
are represented by four-term collision integrals. These four terms in particular case of double Compton scattering with corresponding quantum symmetrization multipliers were for the first time discussed by Chluba \cite{ChlubaThesis}. In previous works this fact was ignored and only two terms were considered in collision integrals, see e.g. \cite{1981ApJ...244..392L, 1981MNRAS.194..439T}.
It should be noted that some properties of plasma, e.g. the detailed balance conditions, may be studied based on only two terms in collision integrals. However, the structure of all four coefficients is different, and their presence in collision integral (\ref{4termsBoltz}) is essential.

In direct triple interactions transition function $W$ can be expressed through differential cross-section
$d\sigma$. Using the definition of $d\sigma$ \cite[Eq.~(64.18)]{1982els..book.....B} and its relation to number of
interactions $dN/dV dt$ per unit volume per unit time from equation~(12.7) of~\cite{landau1975classical}, we
arrive to
\begin{multline}
\label{transfun}
    W_{(1,2|3,4,5)}
    d^3\mathbf{p}_3 d^3\mathbf{p}_4 d^3\mathbf{p}_5 \\
    = c\frac{\sqrt{[\varepsilon_1\varepsilon_2-(\mathbf{p}_1\cdot\mathbf{p}_2) c^2]^2
        -(m_I m_{II} c^4)^2}}
    {\varepsilon_1\varepsilon_2}d\sigma.
\end{multline}
The differential cross-section in turn is related to the QED matrix element squared, averaged over incoming particle
polarizations and summed over outgoing particle polarizations $X$, see~\cite[Eq.~(11.31)]{1976tper.book.....J}.
Therefore
\begin{multline}
    W_{(1,2|3,4,5)} =
    \frac{\alpha r_e^2}{(4\pi)^2}\frac{c^7 X}{      \varepsilon_1\varepsilon_2\varepsilon_3\varepsilon_4\varepsilon_5} \\ \times
    \delta(\varepsilon_{initial}-\varepsilon_{f\!inal})
    \delta^3(\mathbf{p}_{initial}-\mathbf{p}_{f\!inal}),
\end{multline}
where $r_e=e^2/m_e c^2$ is the classical electron radius and $\alpha$ is the fine structure constant. For double Compton scattering $X$ is given by
equations (3), (9), (10) of \cite{1952RSPSA.215..497M}. For relativistic bremsstrahlung $X=16A$, where $A$ is given in
Appendix B of \cite{2004epb..book.....H}.

For all other triple processes listed in Table~\ref{binpint} the matrix elements can be obtained from two aforementioned ones applying the substitution rule, see e.g. \cite[Sec.~8.5]{1976tper.book.....J}. For example, exchanging incoming photon with outgoing electron or positron in double Compton scattering $$e^-_1+\gamma_2 \longrightarrow e^-_3+\gamma_4+\gamma_5$$ the three-photon pair creation-annihilation $$e^-_1+e^+_3 \longrightarrow \gamma_2+\gamma_4+\gamma_5$$ process is recovered. Then the matrix element $X$ of this process can be derived from $X$ of the one of the double Compton scattering by the following substitutions
\begin{equation}
    \mathbf p_3 \longrightarrow -\mathbf p_3,\quad
    \varepsilon_3 \longrightarrow -\varepsilon_3,\quad
    \mathbf p_2 \longrightarrow -\mathbf p_2,\quad
    \varepsilon_2 \longrightarrow -\varepsilon_2.
\end{equation}

In order to get collision integrals for inverse triple interactions we can make use of detailed equilibrium conditions,
valid for QED interactions \cite{1973rela.conf....1E,groot1980relativistic}, that gives
\begin{equation}
    \frac{h^3}{g_{III}}\frac{h^3}{g_{IV}}\frac{h^3}{g_{V}}W_{(1,2|3,4,5)}=
    \frac{h^3}{g_{I}}\frac{h^3}{g_{II}}W_{(3,4,5|1,2)},
\end{equation}
and since in our case all $g=2$ we finally obtain
\begin{equation}
    h^3W_{(1,2|3,4,5)}=2W_{(3,4,5|1,2)},
\end{equation}
The collision rate is a multidimensional integral over the phase space of interacting particles, which spans $4\times 3=12$ dimensions for binary interactions and $5\times3=15$ dimensions for triple ones. The energy-momentum conservation allows to perform integration over 4 variables, while spherical symmetry in the phase space allows to perform additional three integrals, leaving 5 and 8 integrals for binary and triple interactions, correspondingly. From the computational viewpoint this problem is highly demanding. In the next section we introduce the fast numerical scheme dealing with it.

\section{Numerical treatment of the collision integrals}

The main difference between the approach to the kinetics of plasma with Boltzmann equations
\cite{2007PhRvL..99l5003A, 2009PhRvD..79d3008A, 2010PhRvE..81d6401A} and the present approach with Uehling-Uhlenbeck equations
appears in the dependence of emission and absorption coefficients not only on the distributions of the incoming
particles, but the on distributions of the outgoing particles as well, owing to the presence of Bose enhancement and
Pauli blocking factors (further referred to as quantum corrections). Due to this difference in what follows we adopt a ``process-oriented'' technique \cite{2018JCoPh.373..533P, 2019PhLA..383..306P}.

Due to the spherical symmetry in the phase space\footnote{Alternatively, cylindrical coordinates can be used, if anisotropy is present in the problem, see e.g. \cite{BENEDETTI2013206}.}, we introduce spherical coordinates and discretize it. Zone $\Omega^I_{a,j,k}$ of particle kind $I$ corresponds to energy $\varepsilon_a$, cosine of polar angle $\mu_j$ and azimuthal angle $\phi_k$, where indices span the ranges $1\leq a \leq n_\varepsilon$, $1\leq j \leq n_\mu$, $1\leq k \leq n_\phi$. Zone edges are
$\varepsilon_{a\mp1/2}$, $\mu_{j\mp1/2}$, $\phi_{k\mp1/2}$. Width of $a$-th energy zone $\Omega^I_{a}$ is equal to
$\Delta\varepsilon_{a} \equiv \varepsilon_{a+1/2} - \varepsilon_{a-1/2}$. Due to isotropy $f_I$ does not depend on
$\mu$ and $\phi$, therefore particle density $I$ in the zone $a$ is
\begin{multline}
\label{Ydef}
Y^I_{a}(t)=4\pi\int_{\varepsilon_{a-1/2}}^{\varepsilon_{a+1/2}}
c^{-3}\varepsilon\sqrt{\varepsilon^2-m_I^2c^4}
f_I(\varepsilon,t)d\varepsilon \\
\approx 4\pi c^{-3}\varepsilon_a\sqrt{\varepsilon_a^2-m_I^2c^4}
f_I(\varepsilon_a,t)\Delta\varepsilon_a.
\end{multline}
In these variables discretized Uehling-Uhlenbeck equation for particle $I$ and energy zone $a$ is
\begin{equation}\label{discrBoltzmann}
    \frac{d Y^{I}_a(t)}{dt}=
    \sum\left[\eta^{I}_{a}(t)-\chi^{I}_{a}(t)Y^{I}_{a}(t)\right],
\end{equation}
where the sum is taken over all processes that include particle $I$. Emission and absorption coefficients on grid are
obtained by integration of collision integrals over the zones, and these integrals are replaced by sums over
the grid.


The exact conservation laws (particle number, energy, charge) are satisfied in the code thanks to implementation of the interpolation procedure for outgoing particles. The system under consideration has several characteristic timescales for different processes, and the resulting system of ordinary differential equations~(\ref{discrBoltzmann}) is stiff. We use Gear's method
\cite{1976oup..book.....H} to integrate the system numerically.

\subsection{Binary interactions}

First we recall the treatment of binary interactions introduced in \cite{2018JCoPh.373..533P}. Spherical symmetry allows to perform three integrals over the angles of the first incoming particle and over the azimuthal angle of the second one: $\int d\mu_{I} d\phi_{I} d\phi_{II} \longrightarrow 8\pi^2$. $P$-symmetry of QED allows to reduce azimuthal angle range for the third particle $\int_{0}^{2\pi}d\phi_{III} \longrightarrow 2\int_{0}^{\pi}d\phi_{III}$. In addition, energy-momentum conservation expressed with the $\delta$-function in the transition function (\ref{transfun}) allows to perform additional four integrations, with the usual choice to exclude energy and angles of the first outgoing particle and energy
of the second outgoing particle
\begin{multline}
\label{deltatrad}
    \delta(\varepsilon_{1}+\varepsilon_{2}-\varepsilon_{3}-\varepsilon_{4})
    \delta^3(\mathbf{p}_{1}+\mathbf{p}_{2}-\mathbf{p}_{3}-\mathbf{p}_{4})=\\
    c^{2}\frac{\delta(\varepsilon_{4}-\varepsilon_{4}^*)\delta(\varepsilon_{3}-\varepsilon_{3}^*)
    \delta(\varphi_{3}-\varphi_{3}^*)\delta(\mu_{3}-\mu_{3}^*)}
    {\varepsilon_{3}p_{3}\,[1-
    (\beta_{3}/\beta_{4})\mathbf n_{3}\cdot\mathbf n_{4}]},
\end{multline}
where $\beta=pc/\varepsilon$, asterisks denote values defined for excluded integration variables by the energy and momentum conservation
\begin{gather}
    p^*_4=\frac{AB\pm\sqrt{A^2+4m_{IV}^2c^2(B^2-1)}}{2(B^2-1)},\label{Kinematics}\\
    A=\frac{c}{\varepsilon}\left[p^2+(m_{III}^2-m_{IV}^2)c^2\right]
    -\frac{\varepsilon}{c},\qquad
    B=\frac{c}{\varepsilon}\mathbf n_4\cdot{\mathbf p},\nonumber\\
    \varepsilon^*_4=c\sqrt{(p^*_4)^2+m_{IV}^2c^2},\qquad 
    \varepsilon^*_3=\varepsilon-\varepsilon^*_4,\qquad
    \mathbf p^*_3={\mathbf p}-\mathbf p^*_4,\nonumber\\
    \mathbf n_i=\mathbf p_i/p_i,\quad 
    \varepsilon=\varepsilon_1+\varepsilon_2,\quad
    {\mathbf p}=\mathbf p_1+\mathbf p_2.\nonumber
    \label{KinematicsEnd}
\end{gather}

Then the absorption coefficient for incoming particle $I$ in binary interaction $I+II\rightarrow III+IV$ can be
written as
\begin{multline}\label{YrateI}
    \chi^{I}_{a}(t)Y^{I}_{a}(t)\approx
    \frac{\hbar^2c^{4}}{8(4\pi)^2}\sum
    \Delta\mu_{II}\Delta\mu_{IV} \Delta\phi_{IV} \times | M_{fi}|^2
    \\* \times
    \frac{p_4}
    {\varepsilon_{3}[1-(\beta_{3}/\beta_{4})\mathbf n_{3}\cdot\mathbf n_{4}]}
    \times
    \frac{Y^I_{a}(t)}{\varepsilon_a}\frac{Y^{II}_{b}(t)}{\varepsilon_b} \\
    \times\left[1\pm \frac{Y^{III}_{c'}(t)}{\bar Y^{III}_{c'}}\right]
    \left[1\pm \frac{Y^{IV}_{d'}(t)}{\bar Y^{IV}_{d'}}\right],
\end{multline}
where $\bar Y^I_{a}$ are defined as in eq. (\ref{Ydef}) with the substitution $f\rightarrow 2/h^3$ and 
index primes meaning is as follows.

The energies of incoming particles are fixed in the nodes of the grid, but the energies of outgoing particles are off the grid. In order to satisfy conservation laws (particle number, energy, charge) an interpolation procedure is implemented for outgoing particles. Each outgoing particle is split in two interpolated particles between the adjacent zones with a weight determined by its energy and the zone width\footnote{In order to satisfy exactly momentum conservation additional splitting is needed in angular grid \cite{BENEDETTI2013206}.}. Interpolation redistributes outgoing particle $I$ of energy $\varepsilon_a$ into two energy zones $\Omega^{I}_{n}$, $\Omega^{I}_{n+1}$ defined by $\varepsilon_{n} < \varepsilon_a < \varepsilon_{n+1}$, see Eq.~(\ref{csol}) below.

Interpolation procedure for outgoing particles should satisfy the laws of quantum statistics. Any process where a fermion end up in a fully occupied zone is not allowed. We introduce therefore Bose enhancement and Pauli blocking coefficients as
\begin{gather}
    \left[1\pm \frac{Y^{I}_{a'}(t)}{\bar Y^{I}_{a'}}\right]=
    \min\left(1\pm \frac{Y^{I}_{n}(t)}{\bar Y^{I}_{n}},
    1\pm \frac{Y^{I}_{n+1}(t)}{\bar Y^{I}_{n+1}}\right).
\end{gather}
This procedure allows us to fulfil quantum statistics requirements by the cost of partly reducing grid resolution.

Hence emission coefficient for outgoing particle~$I$ in inverse process is
\begin{multline}\label{YrateII}
    \eta^{I}_{a}(t)\approx
    \frac{\hbar^2c^{4}}{8(4\pi)^2}\sum
    C_a(\varepsilon_1)
    \Delta\mu_{IV}\Delta\mu_{II} \Delta\phi_{II} \times | M_{fi}|^2
    \\ \times
    \frac{p_2}
    {\varepsilon_{1}[1-(\beta_{1}/\beta_{2})\mathbf n_{1}\cdot\mathbf n_{2}]}
        \times
    \frac{Y^{III}_{c}(t)}{\varepsilon_c}\frac{Y^{IV}_{d}(t)}{\varepsilon_d} \\
    \times\left[1\pm \frac{Y^{I}_{a'}(t)}{\bar Y^{I}_{a'}}\right]
    \left[1\pm \frac{Y^{II}_{b'}(t)}{\bar Y^{II}_{b'}}\right],
\end{multline}
where interpolation coefficients are
\begin{equation}\label{csol}
    C_a(\varepsilon_1)=
    \begin{cases}
        \dfrac{\varepsilon_{a}-\varepsilon_1}{
            \varepsilon_{a}-\varepsilon_{a-1}}, &
        \varepsilon_{a-1} < \varepsilon_1 < \varepsilon_{a},\\[2.5ex]
        \dfrac{\varepsilon_{a+1}-\varepsilon_1}{
            \varepsilon_{a+1}-\varepsilon_{a}}, &
        \varepsilon_{a} < \varepsilon_1 < \varepsilon_{a+1},\\[2.5ex]
        0, & \text{other }\varepsilon_1.
    \end{cases}
\end{equation}

Integration over the angles can be made only once at the beginning of calculations \cite{2009PhRvD..79d3008A,vereshchagin2017relativistic}, and we refer to this phase as integral coefficients calculation. Then we store corresponding integral coefficients for each set of incoming and outgoing particles in the form of three numbers $K_{1,2,3}$ giving for elementary terms in binary interaction
$I+II\rightarrow III+IV$ the following contributions
\begin{gather}\label{elem2p}
\begin{split}
    \dot Y_a(t) &= \dot Y_b(t) = -K_1^{abcd} D,\\
    \dot Y_c(t) &= K_2^{abcd} D,\quad \dot Y_{c+1}(t) = (K_1^{abcd}-K_2^{abcd}) D,\\
    \dot Y_d(t) &= K_3^{abcd} D,\quad \dot Y_{d+1}(t) = (K_1^{abcd}-K_3^{abcd}) D,\\
    D ={}& Y_{a}(t) Y_{b}(t) \left[1\pm \frac{Y_{c'}(t)}{\bar Y_{c'}}\right]
    \left[1\pm \frac{Y_{d'}(t)}{\bar Y_{d'}}\right],
\end{split}
\end{gather}
that fulfil conservation laws by construction. Sum of all contributions from different elementary reactions gives
total collision integrals on the grid with 2-nd order accuracy. As each stored elementary reaction gives 6 terms for the rates $\dot Y$ and we span elementary reaction space, and not $\dot Y$ themselves, we denote such an approach as ``process-oriented''. Looping
through $\dot Y$ is a more traditional approach \cite{2009PhRvD..79d3008A}, but it demands 6-time separate treatment of the same $D$ that slows down
the calculations.

\subsection{Triple interactions}

The scheme is straightforwardly generalized to triple interactions. However, in contrast to the binary case where
index pairs $I, II$ and $III, IV$ can be swapped without any change in the structure of elementary
terms~(\ref{elem2p}), in triple case such a symmetry is absent. So emission and absorption coefficients for the
processes~(\ref{3pdir}) and (\ref{3pinv}) are discussed separately below. Further, standard choice of excluded
$\delta$-function integrations is not very convenient in this case as it leads to substantial slowdown of calculations
due to the branching of solutions in algebraic equations (\ref{Kinematics}) which requires execution of logical commands.

\begin{figure}[ht]
	\centering
		\includegraphics[width=\columnwidth]{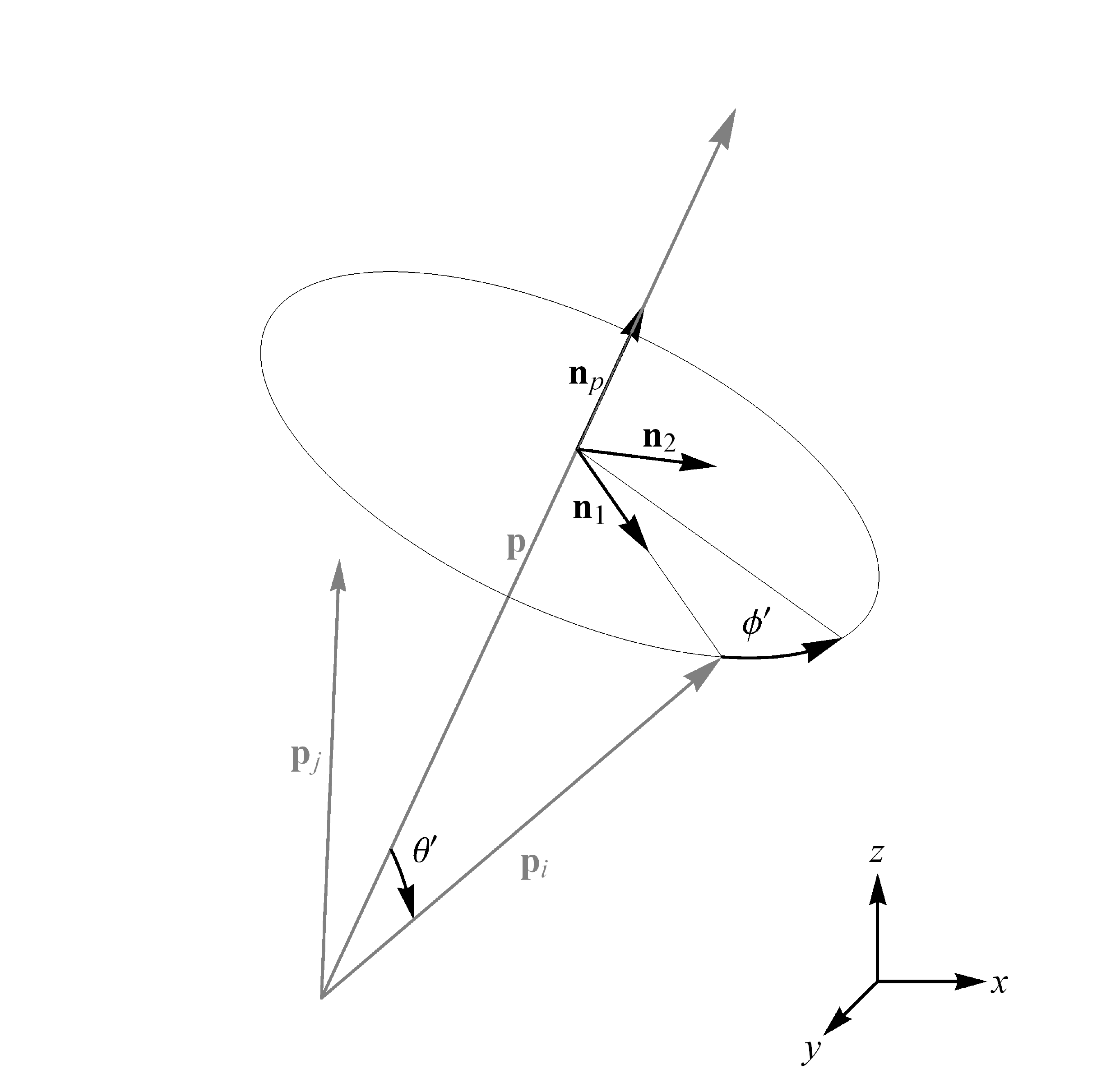}
	\caption{The new coordinates adopted for numerical integration on the phase space.}
	\label{new_kinematics}
\end{figure}
In order to avoid this branching it is possible to use the variables of outgoing particle pair:  a) the energy of one
of these particles $\varepsilon_{i}$, b) its angle $\phi'_{i}$ defined as rotation angle of its momentum $\mathbf p_i$ around the direction of total pair momentum $\mathbf p = \mathbf p_{i} + \mathbf p_{j}$. This azimuthal angle $\phi'_{i}$ serves as an angle of
auxiliary spherical coordinate system $(\mu', \phi')$ with $z'$ axis aligned with $\mathbf p$, see Fig. \ref{new_kinematics}, so that the solution of energy and momentum conservation equations is unique
\begin{gather}\label{Kin}
    \varepsilon_{j} = \varepsilon - \varepsilon_{i}, \quad \phi'_j=\pi+\phi'_i,\\
    \mu'_i=\frac{p^2 + p_i^2 - p_j^2}{2 p p_i},\quad \mu'_j=\frac{p^2 + p_j^2 - p_i^2}{2 p p_j},\nonumber\\
    \mathbf p_i=p_i\left(\mu'_i \mathbf n_p+\sqrt{1-\mu'_i{}^2}
    \left[\mathbf n_1\cos\phi'_i + \mathbf n_2\sin\phi'_i\right]\right),\nonumber\\
    \mathbf p_j=p_j\left(\mu'_j \mathbf n_p + \sqrt{1-\mu'_j{}^2}
    \left[\mathbf n_1\cos\phi'_j + \mathbf n_2\sin\phi'_j\right]\right),\nonumber\\
    \mathbf n_p = \mathbf p/p,\quad p=\sqrt{\mathbf p\cdot\mathbf p},\quad \varepsilon = \varepsilon_{i} + \varepsilon_{j}, \nonumber\\
     p_i = \sqrt{\varepsilon_i^2/c^2-m_i^2c^2}, \quad
    p_j=\sqrt{\varepsilon_j^2/c^2-m_j^2c^2},\nonumber
\end{gather}
where $\mathbf n_1$ and $\mathbf n_2$ are two orthogonal unit normals to $\mathbf p$. The limits of integration are very simple
\begin{gather}\label{IntLimits}
    \phi'_{i} \in [0,2\pi),\qquad \varepsilon_{i} \in [A-B, A+B],\\
    A = \frac{\varepsilon \left(\varepsilon^2-p^2c^2-m_{j}^2c^4+m_{i}^2c^4\right)}
    {2 \left(\varepsilon^2-p^2c^2\right)},\nonumber\\
    B = \frac{p c \sqrt{\left(\varepsilon^2-p^2c^2-m_{j}^2c^4-m_{i}^2c^4\right)^2-(2 m_{i} m_{j} c^4)^2}}
    {2 \left(\varepsilon^2-p^2c^2\right)},\nonumber
\end{gather}
and $\delta$-function becomes
\begin{multline}\label{Jacob}
    \delta(\varepsilon-\varepsilon_{i}-\varepsilon_{j})
    \delta^3(\mathbf p-\mathbf{p}_{i}-\mathbf{p}_{j})=
    \\ \frac{\delta(\mu_{i}-\mu_{i}^*)\delta(\varepsilon_{j}-\varepsilon_{j}^*)\delta(\mu'_{j}
        -\mu'_{j}{}^*)\delta(\phi'_{j}-\phi'_{j}{}^*)}{p p_i p_j},
\end{multline}
where asterisks denote solution of kinematic equations (\ref{Kin}).

The Jacobian~(\ref{Jacob}) exhibits kinematic singularity with a clear physical origin: it appears when total or any particle momentum is zero.
Singularities at $p_i=0$ and $p_j=0$ disappear at the collision integral as the corresponding rates tend to a constant due to
velocity dependence. The only survived singularity is of zero total momentum $p=0$: this turns out to be
center-of-mass frame for the chosen particles and in this case the interval of $\varepsilon_{i}$ in (\ref{IntLimits})
reduces to a point $\varepsilon_{i}=A$, while $\mu'_i$ and $\phi'_i$ span all the sphere. As a result, this
singularity in the integrand is compensated by shrinking of the integration limits of $\varepsilon_{i}$ and the integral remains finite at this point.
However, the subspace of $p=0$ is of much lower dimension than the total phase space of integration (it is 3-dimensional one
versus 6-dimensional one in our case), so this singularity problem can be avoided numerically just by rotating slightly each
of the $\{\mu, \phi\}$ grids for the particles in the reaction to exclude calculations at the $p=0$ points.

Additional advantage of this scheme is the ease of treatment of phase space symmetries: to exclude the same particle
states from incomes and/or outcomes it is sufficient to use ascending (or descending) energy requirements for the same
particles, i.e. in three-photon annihilation we just integrate only over
$\varepsilon_{\gamma_1}<\varepsilon_{\gamma_2}<\varepsilon_{\gamma_3}$. In the traditional approach
$\varepsilon_{\gamma_2}$ and $\varepsilon_{\gamma_3}$ are involved functions of angles of integration, that leads to
another branching in the innermost integration calculations, or alternatively to calculation over all the phase space for
photons (effectively increasing the amount of calculations 6-fold) and introduction of 1/6 multipliers after.

Finally, for the direct process~(\ref{3pdir}) we arrive to the grid representation of the absorption coefficient for
particle~$I$ in the following simple form
\begin{multline}\label{Yrate3I}
    \chi^{I}_{a}(t)Y^{I}_{a}(t)\approx
    \frac{\alpha c r_e^2}{32 \pi ^2}
    \sum\Delta \mu_{II} \Delta \varepsilon_{III}\Delta \mu_{III}\Delta \phi_{III}\Delta \varepsilon_{IV}\Delta \phi'_{IV} \\
    \times \frac{p_{III}}p X 
    \frac{Y^I_{a}(t)}{\varepsilon_a}\frac{Y^{II}_{b}(t)}{\varepsilon_b} \\
    \times \left[1\pm \frac{Y^{III}_{c'}(t)}{\bar Y^{III}_{c'}}\right]
    \left[1\pm \frac{Y^{IV}_{d'}(t)}{\bar Y^{IV}_{d'}}\right]
    \left[1\pm \frac{Y^{V}_{f'}(t)}{\bar Y^{V}_{f'}}\right],
\end{multline}
while for the emission coefficient of particle~$IV$ we have
\begin{multline}\label{Yrate3Ia}
    \eta^{IV}_{d}(t)\approx
    \frac{\alpha c r_e^2}{32 \pi ^2}
    \sum\Delta \mu_{II} \Delta \varepsilon_{III}\Delta \mu_{III}\Delta \phi_{III}\Delta \varepsilon_{IV}\Delta \phi'_{IV} \\
    \times \frac{p_{III}}p X  C_d(\varepsilon_{IV})
    \frac{Y^I_{a}(t)}{\varepsilon_a}\frac{Y^{II}_{b}(t)}{\varepsilon_b} \\
    \times \left[1\pm \frac{Y^{III}_{c'}(t)}{\bar Y^{III}_{c'}}\right]
    \left[1\pm \frac{Y^{IV}_{d'}(t)}{\bar Y^{IV}_{d'}}\right]
    \left[1\pm \frac{Y^{V}_{f'}(t)}{\bar Y^{V}_{f'}}\right],
\end{multline}
the interpolation coefficient $C(\varepsilon)$ being defined in eq.~(\ref{csol})
and primed indices having the meaning described after eq.~(\ref{YrateI}).
We use also special refined grid for $\varepsilon_{IV}$, that ensures the given number of integration points for each
possible zone set of $IV$ and $V$ particles (typically 2 points, further refinement does not change results
substantially). Analogous relations hold for the inverse process.

For each process the sum over the angles again can be performed only once, at the beginning of calculations. Elementary reaction now is represented by three numbers $K_{1,2,3}$ giving the following collision
integral terms
\begin{gather}\label{elem3pdir}
\begin{split}
    \dot Y_a(t) &= \dot Y_b(t) = -\dot Y_c(t) = -K_1^{abcd\!f} D,\\
    \dot Y_d(t) &= K_2^{abcdf} D,\quad \dot Y_{d+1}(t) = (K_1^{abcdf}-K_2^{abcdf}) D,\\
    \dot Y_f(t) &= K_3^{abcdf} D,\quad \dot Y_{f+1}(t) = (K_1^{abcdf}-K_3^{abcdf}) D,\\
    D ={}& Y_{a}(t) Y_{b}(t) \left[1\pm \frac{Y_{c'}(t)}{\bar Y_{c'}}\right]
    \left[1\pm \frac{Y_{d'}(t)}{\bar Y_{d'}}\right] \left[1\pm \frac{Y_{f'}(t)}{\bar Y_{f'}}\right],
\end{split}
\end{gather}
for direct reaction and
\begin{gather}\label{elem3pinv}
\begin{split}
    \dot Y_a(t) &= \dot Y_b(t) = \dot Y_c(t) = -K_1^{abcd\!f} D,\\
    \dot Y_d(t) &= K_2^{abcdf} D,\quad \dot Y_{d+1}(t) = (K_1^{abcdf}-K_2^{abcdf}) D,\\
    \dot Y_f(t) &= K_3^{abcdf} D,\quad \dot Y_{f+1}(t) = (K_1^{abcdf}-K_3^{abcdf}) D,\\
    D ={}& Y_{a}(t) Y_{b}(t) Y_{c}(t)
    \left[1\pm \frac{Y_{d'}(t)}{\bar Y_{d'}}\right] \left[1\pm \frac{Y_{f'}(t)}{\bar Y_{f'}}\right],
\end{split}
\end{gather}
for inverse reaction. To find rates $\dot Y$ we span elementary reaction space as in the case of binary
interactions.

\section{Numerical results}

In this section the results of numerical calculations are presented. We compare our results for collision integral with all known analytical expressions in the literature. They appear to use only Boltzmann statistics
\begin{equation}
\bar f_{eq}=\exp\left(-\frac{\varepsilon-\mu}{k_BT}\right),
\label{Bstat}
\end{equation}
without quantum corrections, so in what follows all comparisons are done with such a case. Svensson \cite{1984MNRAS.209..175S} gives analytical expressions for thermal photon emission coefficients $\eta_\gamma$ in the soft photon limit ($\varepsilon_\gamma \ll k_B T$) for the double Compton scattering, electron-electron bremsstrahlung, three-photon annihilation and radiative pair production. These rates are inversely proportional to the photon energy $\eta_\gamma \sim \varepsilon^{-1}$. We note that Svensson formula for electron-electron bremsstrahlung does not correctly describe the non-relativistic limit, therefore we use the formula of Haug\cite{1975ZNatA..30.1099H}, which represents non-relativistic limit. One should keep in mind that Svensson formulas represent an interpolation between non-relativistic and ultra-relativistic limits and their accuracy for the intermediate plasma temperatures is not estimated. For this reason we compare our results only in a non-relativistic domain, selecting for $k_B T=0.1 m_e c^2$. Below dimensionless energy $e\equiv\varepsilon/m_e c^2$ and temperature $\theta = k_B T/m_e c^2$ are used. 

The calculations are performed on a logarithmic energy grid with $n_\varepsilon=60$ nodes, with different homogeneous grids for angular variables, $\phi$-grid is 2 time denser then $\mu$-grid (typically $\mu$-grid contains $n_\mu=16$ nodes). In order to resolve better soft photons we choose the lower particle energy boundary $0.001m_e c^2$ and upper particle energy boundary $10m_e c^2$. This energy region perfectly covers thermal distribution of particles from $\theta=0.1$ to $\theta=1$.  

\begin{figure}[tbp]
\centering
\includegraphics[width=\columnwidth]{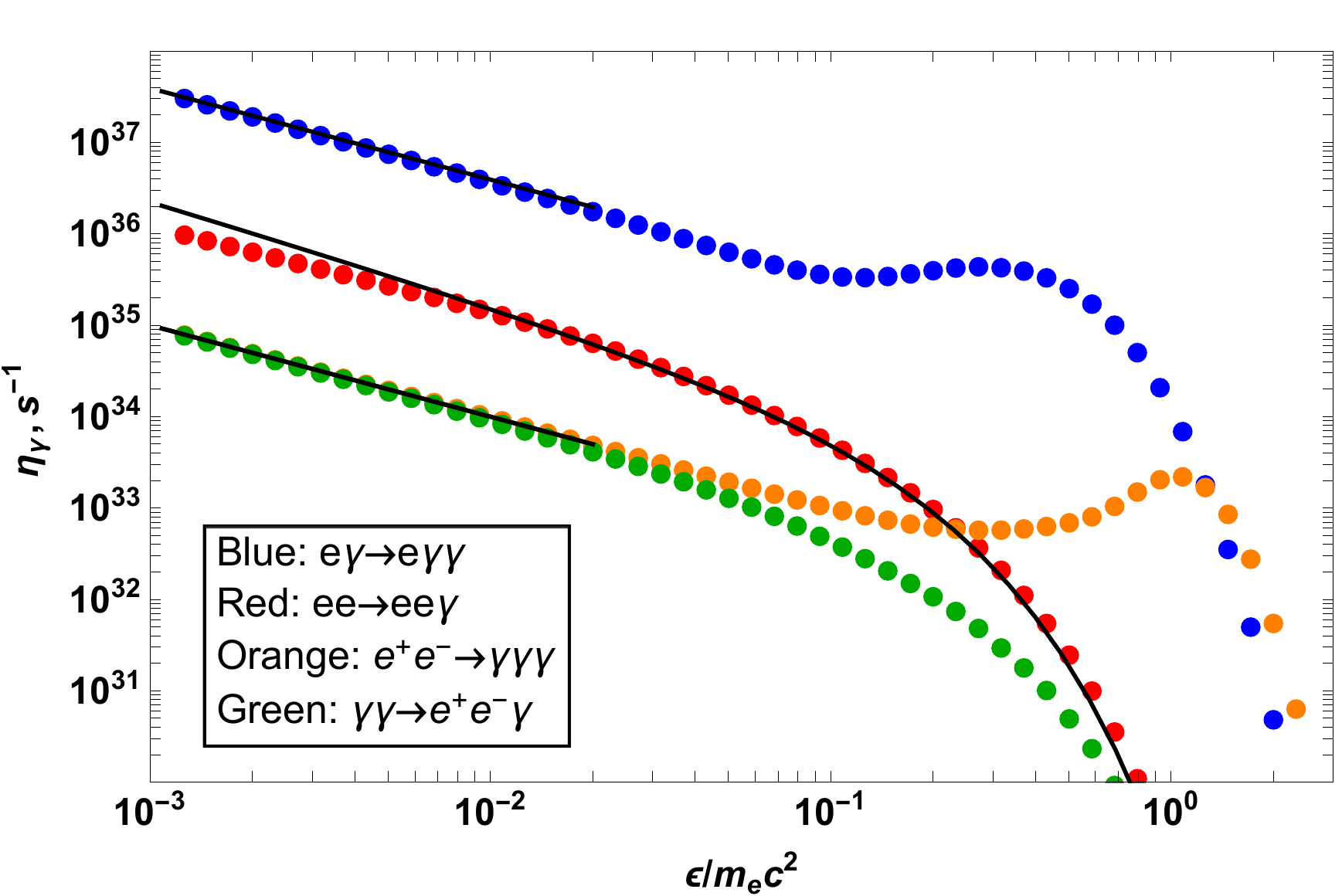}
\caption{Thermal photon emissivity $\eta_\gamma$ as a function of photon energy. Solid curves represent Svensson \cite{1984MNRAS.209..175S} and Haug \cite{1975ZNatA..30.1546H} analytical formulas, see text for details. From top to bottom: double Compton scattering, relativistic bremsstrahlung, three-photon annihilation, radiative pair production. The number of nodes is: 60 for energy, 32 for $\phi$ angle and 16 for $\mu$ angle. The distribution functions are taken at equilibrium with $\theta=0.1$ and chemical potential being zero.}
\label{fig_ph_em_Sv_Ha}
\end{figure}
\begin{table}[tbp]
\caption{Numerical accuracy measured by the $Q$ coefficient for selected number of angular grid nodes ($n_\varepsilon=60, n_\phi=2n_\mu$) for different interactions.
\label{qtab}} \center
\begin{tabular}[c]{|l|l|l|l|l|}
\hline
Process/$n_\mu$         & 4        & 8        & 16         & 32 \\
\hline
Double Compton scattering      & 0.143   & 0.087 & 0.044 & 0.049   \\
\hline
relativistic bremsstrahlung    & 0.396  & 0.336  & 0.271 & 0.228  \\
\hline
three photon annihilation      & 0.043  & 0.012 & 0.023 & 0.021 \\
\hline
radiative pair production      &  0.275 & 0.133 & 0.085 & 0.075   \\
\hline
\end{tabular}
\end{table}

\begin{figure}[tbp]
\centering
\includegraphics[width=\columnwidth]{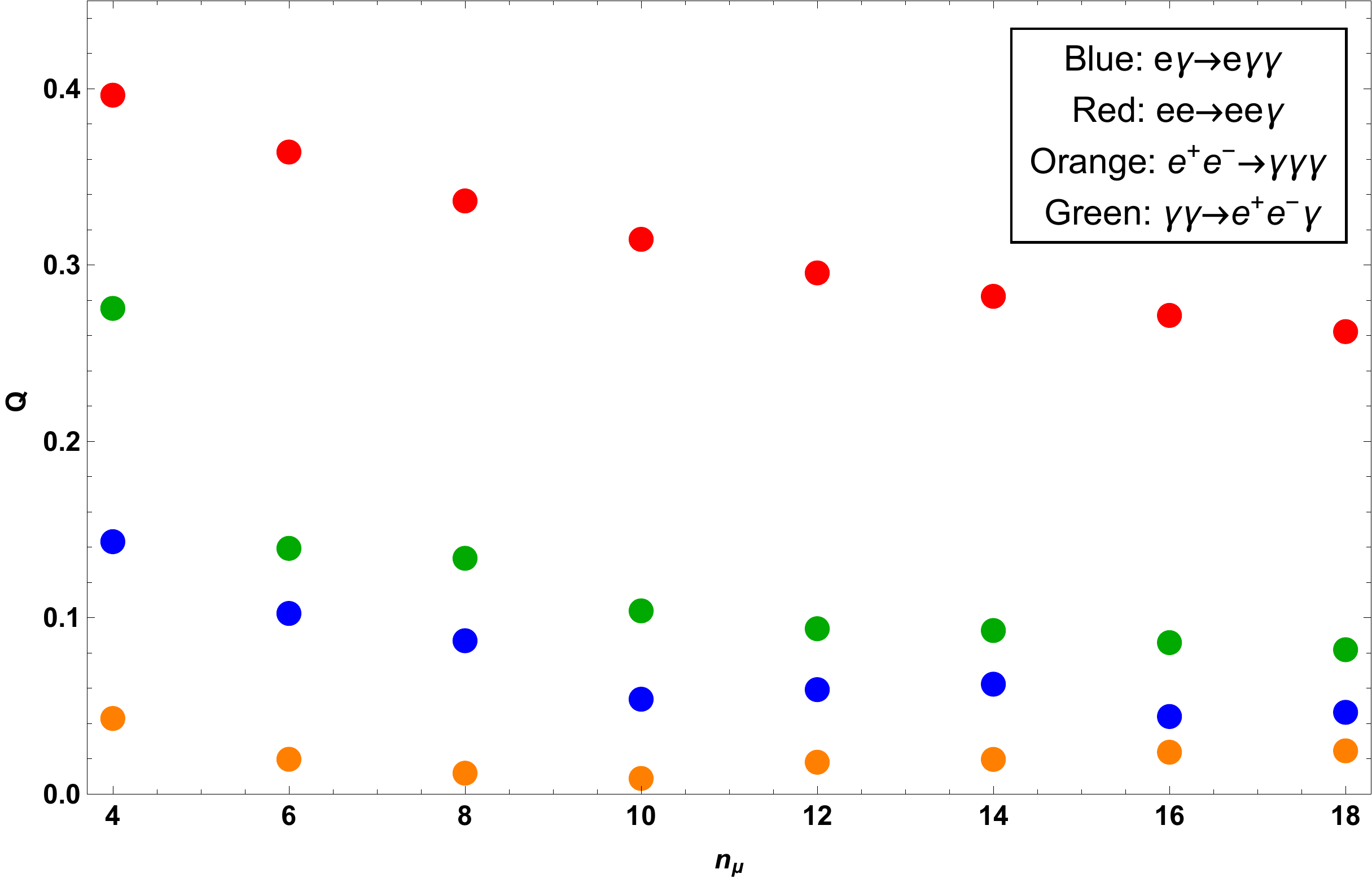}
\caption{Dependence of $Q$ parameter defined in eq. (\ref{Qpar}) which characterize accuracy of the calculation of triple interactions on the number of nodes in angle variables. The accuracy increases with increasing number of nodes.}
\label{fig_qtab}
\end{figure}
Figure \ref{fig_ph_em_Sv_Ha} shows numerical thermal photon emissivity coefficient compared with analytic formulas (black curves). All solid curves besides bremsstrahlung are plotted up to the photon energy $e=0.02$, as they are valid only in the soft photon limit. There is a good agreement with Svensson formula (A10) in \cite{1984MNRAS.209..175S} for the double Compton scattering (blue dots), see also Table \ref{qtab} and Figure \ref{fig_qtab}. Bremsstrahlung emissivity (red dots) show deviations from Haug formula (2.10) in \cite{1975ZNatA..30.1546H} for soft photons; this deviation decreases with increasing resolution in angles, see Table \ref{qtab}. Emissivities for three-photon annihilation (orange dots) and radiative pair production (green dots) are compared with eq. (A18) and (A20) in \cite{1984MNRAS.209..175S}, respectively, differ by a factor of particle density in the case of non-zero chemical potential and become identical in the case of zero chemical potential. The latter case is presented in Figure \ref{fig_ph_em_Sv_Ha}. Indeed, thermal emissivities of both processes coincide for soft photons, and become different with increasing photon energy. Overall, our numerical results show a good agreement with the corresponding non-relativistic formulas.

In order to estimate accuracy of our calculations, we introduce the following quantity for each process
\begin{equation}
Q=n^{-1}_\text{cut}\sum_{a} \left|\frac{\eta_{a}}{\eta_\gamma(e_a)}-1\right|,
\label{Qpar}
\end{equation}
where $\eta_a$ is given by eq. (\ref{Yrate3Ia}) for Boltzmann statistics. This coefficient expresses the average relative deviation of numerical results from analytical ones for energy grid nodes. Limit of soft photons is adopted to summation index with upper boundary $n_\text{cut}$ (namely $n_\text{cut}=15$ as $e_{15}=0.01=0.1\theta$). Electron-electron bremsstrahlung emissivity is compared over full energy domain. Table \ref{qtab} and Fig. \ref{fig_qtab} present values of $Q$ for selected number of angular grid nodes. The relative error generally decreases with angular grid refinement. The $Q$ value saturates for large number of nodes, which indicates that further energy grid refinement is needed. The relatively large errors for relativistic bremmstrahlung can be explained as follows. As calculations of Haug\cite{2004epb..book.....H} show the differential cross section of bremsstrahlung strongly depends on angles even at nonrelativistic energies. Low resolution in angular grid does not allow to capture this strong dependence, which results in reduced accuracy, being compared to other interactions. Both energy and angular grid refinements provide substantial improvements of accuracy for bremsstahlung, see Fig.~\ref{fig_conv1} and \ref{fig_conv2} below. Overall convergence of numerical results to the corresponding analytical ones is from good to satisfactory.
\begin{figure}[tbph!]
\centering
\includegraphics[width=\columnwidth]{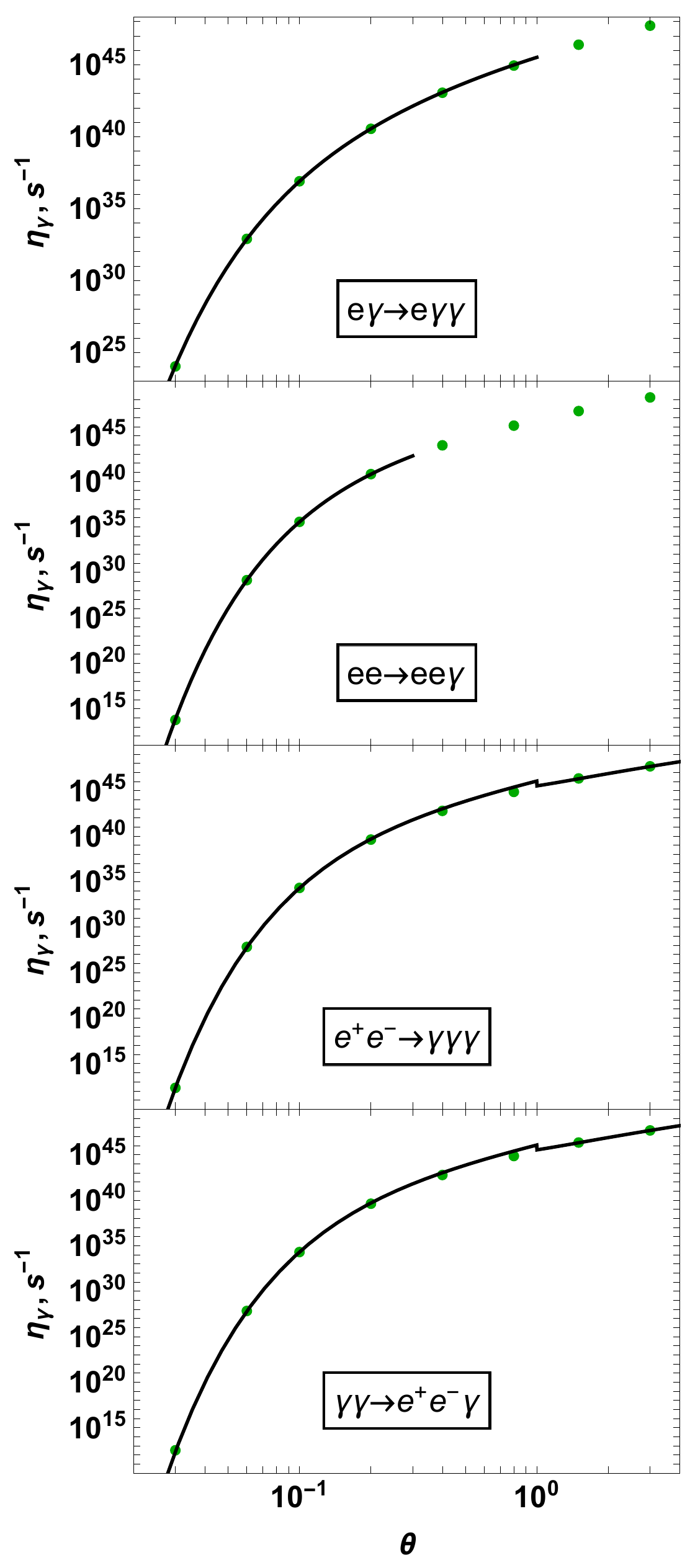}
\caption{Thermal photon emissivity $\eta_\gamma$ as a function of temperature with fixed energy $e=0.05k_B T$. Solid curves represent Svennson \cite{1984MNRAS.209..175S} and Haug \cite{1975ZNatA..30.1546H} analytical formulas, see text for details.}
\label{fig_ph_em_theta}
\end{figure}
\begin{figure}[tbph!]
\includegraphics[width=\columnwidth]{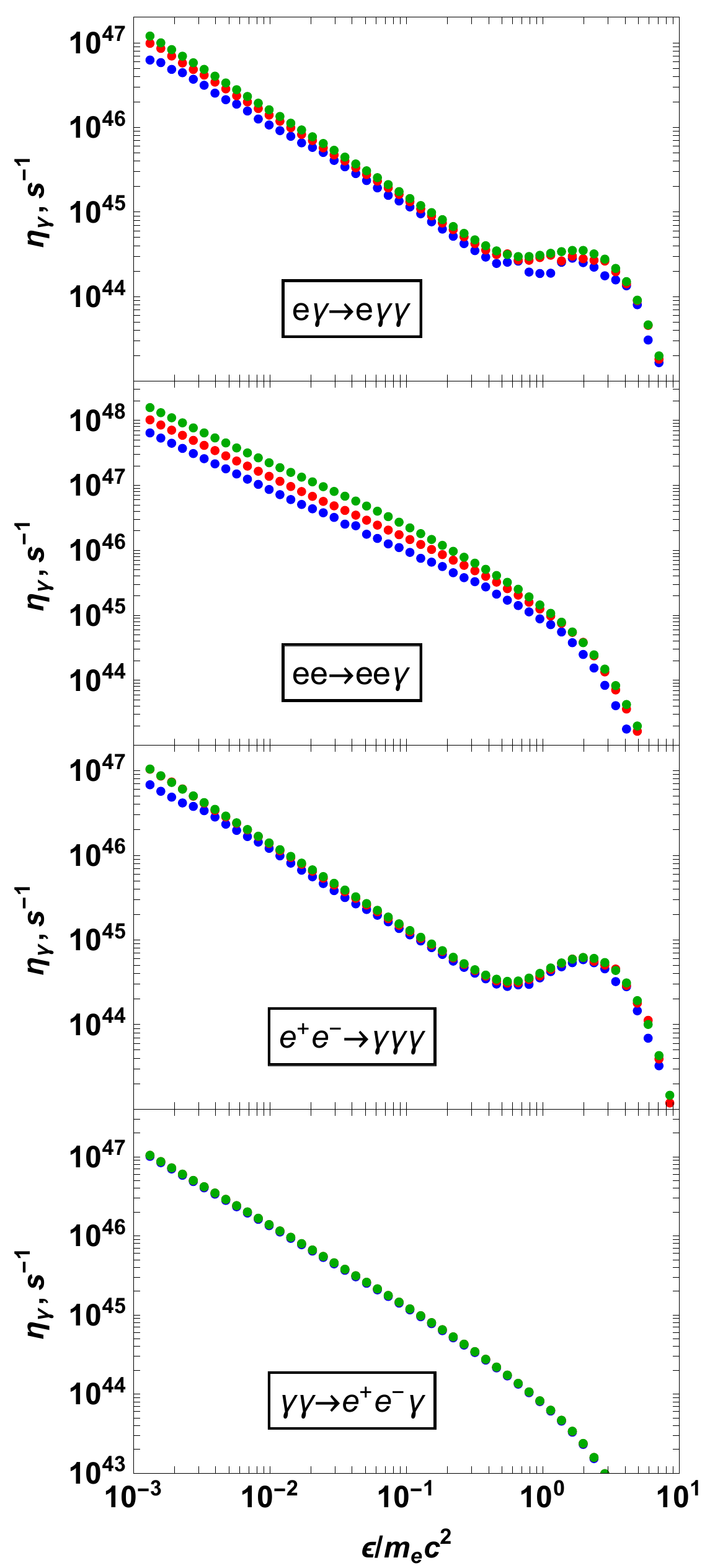}
\caption{Convergence of emission coefficients with increasing grid resolution, for a fixed temperature $\theta=1$. The number of nodes in angles is $4$ (blue dots), $8$ (red dots), $16$ (green dots), the number of energy nodes is $60$. From top to bottom: double Compton scattering, relativistic bremsstrahlung, three-photon annihilation, radiative pair production.}
\label{fig_conv1}
\end{figure}
\begin{figure}[tbph!]
\includegraphics[width=\columnwidth]{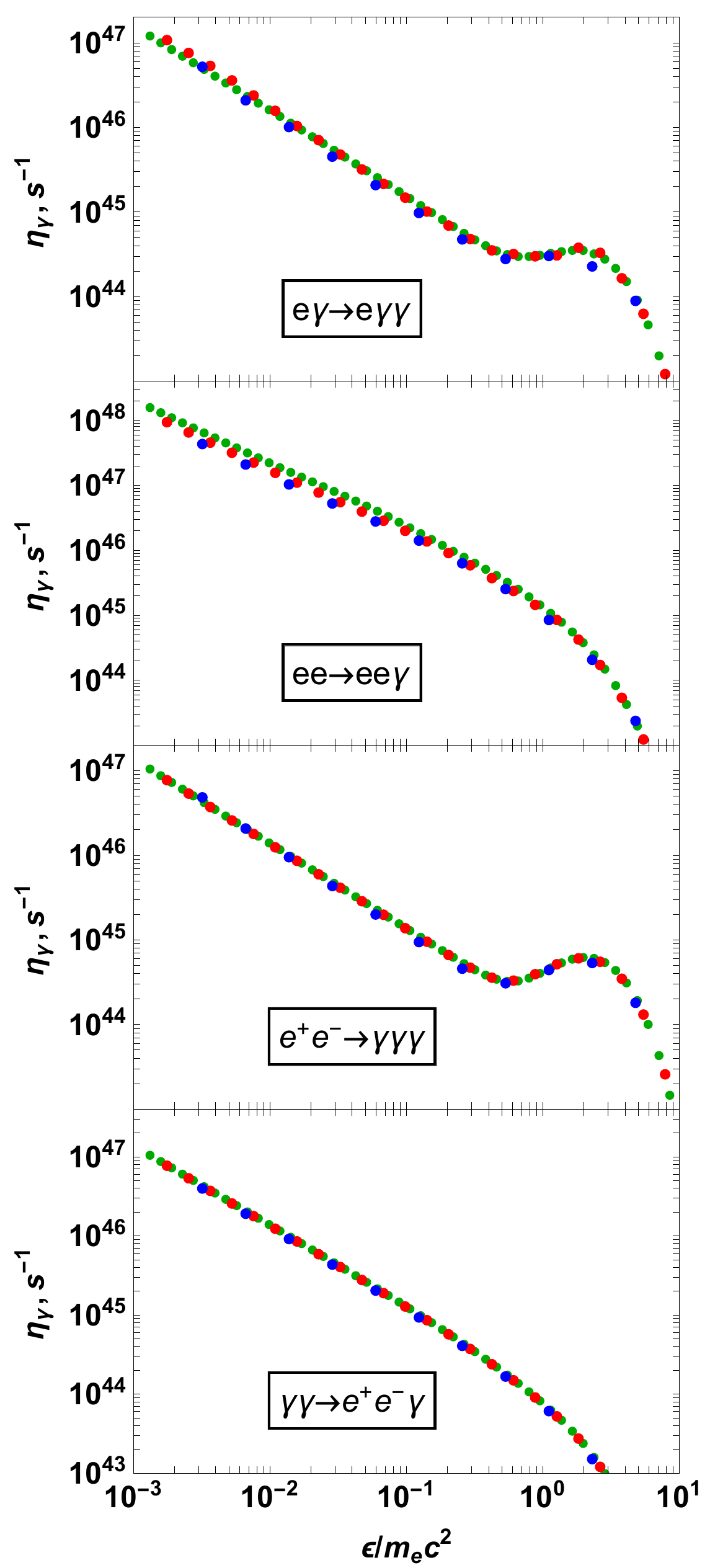}
\caption{Convergence of emission coefficients with increasing grid resolution, for a fixed temperature $\theta=1$. The number of nodes in energy is $15$ (blue dots), $30$ (red dots), $60$ (green dots), the number of nodes in angles is $16$. From top to bottom: double Compton scattering, relativistic bremsstrahlung, three-photon annihilation, radiative pair production.}
\label{fig_conv2}
\end{figure}

We also present the results of calculations in a wide temperature region $\theta=\{ 0.03, 0.06, 0.1, 0.2, 0.4, 0.8, 1.5, 3\}$ which extends from non-relativistic to mildly relativistic temperatures. We provide a comparison with a fixed photon energy in the soft photon limit, here $e=0.05 \theta$. The result is shown in Figure \ref{fig_ph_em_theta}, where black solid curve show analytic curves of Svensson and Haug. The analytical curve for double Compton scattering is plotted up to $\theta=1$, and the non-relativistic analytical curve for bremsstrahlung is plotted up to $\theta=0.3$. Three-photon annihilation and radiative pair production are compared with the same analytic curves (normalized to different number densities), which consist of two segments: non-relativistic one plotted up to $\theta=1$ and relativistic one plotted above $\theta=1$.

The convergence of our results is demonstrated in Figures \ref{fig_conv1} and \ref{fig_conv2}, where we present the dependence of emission coefficients at fixed temperature $\theta=1$ with varying number of nodes in angles (Fig. \ref{fig_conv1}) and in energy (Fig. \ref{fig_conv2}).

\section{Conclusions}

In this work we present the first principle calculations of collision integrals in triple interactions in relativistic plasma, including double Compton scattering, relativistic bremsstrahlung, radiative pair production and three photon creation/annihilation. These processes are important radiative processes in relativistic plasmas, and their account is essential also in the studies of non-equilibrium plasmas. The collision integrals are computed directly by numerical integration of vacuum QED matrix elements over the phase space of interacting particles on the finite grid. The plasma dressing effects such as Debye screening are not considered. Uehling-Uhlenbeck collision integrals computed in this scheme account for quantum statistics of particles, therefore strongly degenerate plasma can be modelled as well. The proposed method allows solution of relativistic kinetic equations for arbitrary non-equilibrium distribution functions, which take into account all triple interactions for the first time. This is an important new step, as most existing kinetic codes do not include triple interactions, and only few include them in a simplified way. In addition, our method can be applied to other processes for which matrix elements are known, such as neutrino interactions.

The comparison with existing analytic results for thermal distributions shows good agreement, with relative errors in the calculations not exceeding few percent, except for the case of relativistic bremsstrahlung where the error can reach up to 20 percent. The convergence of interaction rates with increasing grid resolution is demonstrated.

The new kinetic code, which computes binary and triple interactions in relativistic plasma out of first principles has wide applications in astrophysics, as well as for description of plasmas generated in laboratory.

\acknowledgments
This work is supported within the joint BRFFR-ICRANet-2018 funding programme within the Grant No. F19ICR-001.


\section*{AIP Publishing Data Sharing Policy}
The data that support the findings of this study are available from the corresponding author upon reasonable request.

\bibliography{litera}

\end{document}